\def\gtorder{\mathrel{\raise.3ex\hbox{$>$}\mkern-14mu
		\lower0.6ex\hbox{$\sim$}}}
   \def\ltorder{\mathrel{\raise.3ex\hbox{$<$}\mkern-14mu
		\lower0.6ex\hbox{$\sim$}}}

   \documentstyle[12pt,aas2pp4]{article}
    \begin{document}

   \title{
   An 
   Atlas of {\it Hubble Space Telescope} Ultraviolet Images of Nearby Galaxies$^{1}$}

   \author{Dan Maoz$^{2}$,
   Alexei V. Filippenko$^{3}$,
    Luis C. Ho$^{3,4}$, F. Duccio Macchetto$^{5}$,
    Hans-Walter Rix$^{6}$, and Donald P. Schneider$^{7}$}


   $^1$ Based on observations with the {\it Hubble Space Telescope} which is operated

   by AURA, Inc., under NASA contract NAS 5-26555

   $^2$ School of Physics \& Astronomy and Wise Observatory,

    Tel-Aviv University, Tel-Aviv 69978, Israel. dani@wise.tau.ac.il

   $^3$ Department of Astronomy, University of California, Berkeley, CA 94720-3411

   $^4$ Center for Astrophyics, 60 Garden Street, Cambridge, MA 02138

   $^5$ Space Telescope Science Institute, 3700 San Martin Dr., Baltimore, MD 21218

   $^6$ Steward Observatory, University of Arizona, Tucson, AZ 85721

   $^7$ Department of Astronomy and Astrophysics,

    The Pennsylvania State University, University Park, PA 16802

   \begin{abstract}
\footnotesize
   We present an atlas of UV ($\sim 2300$ \AA) images, obtained with the
   {\it Hubble Space Telescope} ({\it HST}) Faint Object Camera,
    of the central $22''\times 22''$ of
   110 galaxies.
   The observed galaxies are an unbiased selection constituting about one half
  of a complete sample of all large ($D>6'$) and nearby
  ($V< 2000$ km s$^{-1}$) galaxies. This is the first extensive UV imaging survey 
of normal galaxies. The data are useful for studying star formation,
low-level nuclear activity, and UV emission by evolved stellar populations in galaxies.
   At the {\it HST} resolution ($\sim 0.05''$), the images display an 
   assortment of morphologies and UV brightnesses. These include bright
  nuclear point sources, compact young star clusters scattered in the field or
arranged in circumnuclear rings, centrally-peaked diffuse light distributions,
and galaxies with weak or undetected UV emission. We measure the integrated
$\sim 2300$ \AA\ flux in each image, classify the UV morphology, and examine
trends between these parameters and the optical properties of the galaxies.

   \end{abstract}


\footnotesize
   \section{Introduction}

Ultraviolet (UV) imaging is a powerful tool for the study of galaxies. Recent
star formation and low-level nuclear activity can be traced
and measured in UV images, unhindered by the bright background of the older stellar
populations which dominate in visual bands. The source of UV emission in 
some quiescent early-type galaxies can also be studied with UV imaging.
   Prior to the launch of the {\it Hubble Space Telescope} ({\it HST}), few
galaxies were imaged in the UV. These few were mostly Local Group
spiral galaxies observed at low angular resolution using rocket-borne
telescopes on brief flights (see Bohlin et al. 1990, and references therein).
This situation has improved in recent years with the flight of UV imaging telescopes
on the Space Shuttle (e.g., Landsman et al. 1992;
 Hill et al. 1992), although the angular resolution is still low and the
number of galaxies observed is small, due to the limited mission duration.
In this respect, {\it HST}, with its 
high sensitivity and angular resolution, has opened up a new frontier.
Summaries of recent {\it HST} UV-imaging studies of galaxies appear
in Benvenuti, Macchetto, \& Schreier (1996).

   We have carried out a UV imaging survey with the pre-COSTAR
   {\it HST} Faint
   Object Camera (FOC) of the central regions of 110 large nearby 
   galaxies, selected randomly from a complete sample of 240 galaxies.
   The main purposes of the survey are (a) to detect low-luminosity
   active galactic nuclei (AGNs), exploiting the high contrast of a nuclear
   UV continuum source above the low UV background from stars, and
   (b) to study star formation in the central regions of the galaxies; by
   observing at $\sim2300$ \AA, the survey images detect primarily 
   the youngest existing stellar populations and thus provide a clean probe of the
   most recent sites of active star formation, uncontaminated by light
   from more evolved stars. The power of this technique, especially when
   augmented by the high {\it HST} angular resolution, has been recently
   demonstrated in UV studies of starbursts by Conti \& Vacca (1994),
   Meurer et al. (1995), and Maoz et al. (1996).

   Maoz et al. (1995) presented data from the survey for nine galaxies
   displaying bright compact nuclear UV sources, five of which are potential
   low-luminosity AGNs. Maoz et al. (1996) analyzed the images of five
   additional galaxies having circumnuclear star-forming rings, and showed
   that much of the star formation in the rings is confined to massive compact
   clusters that are probably bound.
    In a forthcoming
   work (Ho et al. 1996b) we will use the UV images to
    analyze the star-formation properties of all the galaxies in the sample.
   In this paper, we present images and basic data for the 110 galaxies observed.
   The images provide a reference for the UV brightness and morphology
   of the galaxies in the sample, which can be used as a general guide
   to the UV properties of the centers of galaxies, and as an aid
   in extracting
   the actual data from the {\it HST} archive. We also provide a classification
   of the UV morphology, additional notes on some of the galaxies and what is seen in their
   images, and a measurement of the $\sim 2300$ \AA\ flux integrated over the image.
   We tabulate these together with 
   the larger-scale properties of the galaxies and (for the northern galaxies) the
   spectroscopic nuclear classification from Ho, Filippenko, \& Sargent (1995, 1996a).
   In \S 2, below, we describe the sample selection and the observations.
   In \S 3 we present tables listing the galaxy parameters, a pictorial
   atlas of the data, and notes on some of the individual
   galaxies. In \S 4 we intercompare the UV and optical properties of the galaxies
  and provide a brief summary.

   \section{Observations}

    The sample from which the observed galaxies were chosen consists of all
   galaxies in the UGC and ESO catalogs (Lauberts \& Valentijn 1989)
    with heliocentric velocities available in the literature
 and  less than $2000$ km s$^{-1}$, and photographic diameters (as 
   defined in the catalogs) greater than $6'$.
A total of 22 galaxies were removed
from this initial
sample of 262 galaxies during subsequent stages of the sample
definition. In 
21 of these, no nuclear position
could be defined because they were too diffuse, too low in surface
 brightness, or edge-on with strong dust lanes. 
One galaxy (ESO 1331$-$4517) had a bright foreground 
star near the nucleus, which would have endangered
 the {\it HST} instruments. This left a well-defined sample
of 240 galaxies with observable nuclei.

   Twenty-seven galaxies from this sample were
   not included in our target list
    because they were proposed targets of other {\it HST} UV-imaging programs.
However, we have subsequently obtained from the {\it HST} archive the data for 
   seven of these galaxies which were actually observed with an instrumental configuration
similar to the one we used (FOC + UV filter at $\sim 2300$ \AA). 

 Digitized photographs from the GASP archive at Space Telescope
Science Institute (STScI) of all
 potential target galaxies  were examined
and the coordinates of the nucleus determined to $\sim 1''-2''$ precision
by computing the centroid of the light distribution.
In a few cases,
as evidenced by the eventual {\it HST} images, the galaxy nucleus position
was off by up to $\sim 5''$, due to isophote distortion by dust features 
in the GASP images.

 The observations were done in Snapshot 
mode -- i.e., targets were chosen from the target list by the 
 STScI staff based on the convenience of their location
on the sky. The brief exposure was used to fill the gaps left in the
observing schedule after other science programs had been scheduled.
The observed galaxies are therefore an unbiased selection from the complete sample.

   Among the galaxies described in this paper, 103 were successfully observed with {\it HST}
while the program was active, in
   1993, March through July. We supplement these observations with data for the seven
out of the 27 ``embargoed'' galaxies
that were observed with {\it HST} between 1991 and 1995. We therefore have data for a total
of 110 galaxies out of the complete sample of 240 galaxies. Observations were unsuccessful
   for four galaxies due to wrong coordinates (NGC 660) or
   telescope malfunction (NGC 3714, ESO 0514$-$3709, and NGC 3137);
 we do not count these among the successful 110
   observations, even though data files for them exist in the {\it HST} archive.

   All of our images were obtained with the {\it HST} $f/96$ FOC (Paresce 1990)
   in its ``zoomed'' $1024\times 512$-pixel mode
    with $0.022''\times 0.044''$ pixels,
    giving a field
   of view of $22''\times 22''$. The F220W filter was used. This is
   a broad-band filter with an effective wavelength of $\sim 2300$ \AA\ and
   effective bandpass of $\sim 500$ \AA. 
    The exposure time was 10 minutes per galaxy. We will refer to this configuration
and exposure time
as the ``standard setup.''
   The images were processed by STScI's ``pipeline'' reduction (Baxter
   et al. 1994), after which the pixel scale is $0.0225''$ pixel$^{-1}$. The data
   were obtained before the {\it HST} repair mission at the end of 1993,
   and therefore are affected by spherically aberrated optics. 
   As a result, the point-spread function (PSF) consists of a sharp core
   of full width at half maximum (FWHM) $\sim 0.05''$ that contains about 15\% of the light,
   with the rest of the light
    spread in a complex low-level ``halo'' with a radius of several arcseconds 
   (Burrows et al. 1991).
   In the observing mode we have used, the FOC is limited in its dynamic range
   to 255 counts (8 bits) per zoomed pixel; additional signal causes the
   counts to ``fold over'' and start again from 0. Another problem
   is that the detected count rate becomes nonlinear, gradually saturating
    for bright sources (see Baxter et al. 1994). 
   The central pixels of most of the compact bright sources detected in the 
   images may be in the nonlinear regime, and the brightest of them
   are clearly saturated. In several galaxies where we report the brightness
  and/or angular size of individual compact sources,
 our analysis relies mainly
   on the wings of the PSF, which have low count rates
    ($\ltorder 0.05$ s$^{-1}$ pixel$^{-1}$), using the algorithms described
by Maoz et al. (1995, 1996). We 
   model the PSF  using a well-exposed
   F220W image of a star observed with the FOC $ f/96$ $256\times256$ pixel mode,
   which has a large
   dynamic and linear range (but small field of view). Such empirical PSFs
   are required for work in the UV (Baxter et al. 1994).

   As in Maoz et al. (1995, 1996), we translate the FOC counts to a flux density
   $f_{\lambda}$ at 2270~\AA\ assuming
    1 count s$^{-1}=
   1.66\times 10^{-17}$ erg s$^{-1}$ cm$^{-2}$ \AA$^{-1}$, based on
   the on-line calibration data available from STScI for the FOC and F220W
   filter, with a 25\% increase
   in sensitivity of the $512 \times 1024$ zoomed-pixel 
   mode relative to the $ 512\times 512$ pixel
   mode (Baxter et al. 1994). This calibration assumes a spectrum that is
constant in $f_{\lambda}$. As detailed in Maoz et al. (1995), the F220W count-rate
vs. $f_{\lambda}$ at 2270 \AA\ is  weakly dependent on the spectral slope,
with a change of only a few percent for a large range in slopes.
 The uncertainty in the absolute flux is $\sim 20$\% when measuring
 individual compact sources (Baxter et al. 1994), but can be as small
 as $\sim 5$\% when measuring the UV flux integrated over large areas
 (e.g.\ $\sim 150$ arcsec$^2$; Meurer 1995), if the background can be
 reliably determined.

 An additional concern in UV imaging photometry
 is the presence of ``red leaks'' through the F220W filter, which we define as light
of wavelength $\lambda > 3200$ \AA\ that may pass the low, but non-zero,
transmission of the filter at these wavelengths. This can be
a problem when observing very red sources,
 e.g., the centers of the early-type galaxies
in our sample.
 Tests by the FOC team indicate
that the F220W transmission curve is not significantly changed from that
measured before launch (Baxter et al. 1994), i.e., it peaks
at $\sim 2200$ \AA\ and falls monotonically to the red, decreasing by
a factor of $\sim 1000$ between 2200 \AA\ and 4400 \AA. Using this
transmission curve together with the FOC+{\it HST} efficiency as a function
of wavelength, and the spectral energy distribution templates of early-type
galaxies measured by Kinney et al. (1996), we calculate that a $V$-band 
surface brightness of 13.7 mag arcsec$^{-2}$ produces 
one red-leak count per dezoomed pixel in a 600 s FOC+F220W exposure.
In other words, an (unrealistic) galaxy with {\it no} flux below 3200 \AA\ needs to have
a $V$ surface brightness brighter than 13.7 mag arcsec$^{-2}$ in order to
be detected through the red leak in our standard setup. 

Lauer et al. (1995) have measured with the pre-COSTAR 
{\it HST} the $V$-band surface-brightness
profiles of 45 nearby early-type galaxies. None of their galaxies have
an observed (i.e. before deconvolution of the aberrated PSF) surface brightness
brighter than 13.7 mag arcsec$^{-2}$ within a radius of $0.1''$, so none would have
a detected red-leak signal in our FOC exposures, even within this small 
(4.4 FOC pixels) radius. At a radius of $1''$, where we can still detect a clear
UV signal with the FOC from many of the early-type galaxies in our sample,
Lauer et al. (1995) measure a surface brightness of 15 mag arcsec$^{-2}$
for the brightest galaxies in their sample, and typically 17--18 mag arcsec$^{-2}$.
We conclude that, even without any assumptions about the shape of the spectrum
at wavelengths $\lambda < 3200$~\AA, the red leak contributes negligibly to the
signal we detect from the reddest galaxies over most of the field of view.

 We note that NGC 2997, a spiral galaxy with a
a circumnuclear ring, has an {\it HST} WFPC2 6000 \AA\ image which Maoz et al. (1996) 
have analyzed in conjunction with its FOC 2300 \AA\ image. We have convolved the 6000 \AA\ image, which
is free of spherical aberration, with the pre-COSTAR PSF, and measured the 
surface-brightness profile. The nuclear region, which at a radius of $0.1''$ has a $V$ surface
brightness of 14.6 mag arcsec$^{-2}$, is expected to produce only 0.4 FOC counts pixel $^{-1}$
through the F220W red leak. The nuclear region is, indeed,
 undetected in the UV image, as expected if the
F220W transmission curve has its pre-launch values longward of 3200 \AA.
This comparison also
sets an upper limit, of a factor of 2, by which the F220W transmission curve
could be off from its pre-launch value.
As a further test, we note that two of the galaxies for which Lauer et al. (1995) present
$V$-band profiles, NGC 1023
and NGC 4636, are in our sample as well, and for both we detect with the FOC faint
centrally-peaked diffuse light distributions. In NGC 1023, we measure with
the FOC 2.3 counts pixel$^{-1}$ at the $0.1''$ radius,
 while Lauer et al.'s (1995) observed $V$-band surface brightness at this radius,
 of 14.8 mag arcsec$^{-2}$ (T. Lauer, private communication),
 is similar to that of NGC 2997, which produces no detected F220W counts.
 NGC 4636 has 1/2 as many FOC counts
as NGC 1023 at $0.1''$, even though its observed $V$ surface brightness is 8
times lower at this radius (T. Lauer, private communication).
The FOC counts for both NGC 1023 and NGC 4636 must therefore be dominated
by true UV flux. This is also confirmed for NGC 1023 by the flux at  2300 \AA\ measured
with  the {\it International Ultraviolet Explorer} ({\it IUE}; Kinney et al. 1993), which is
consistent with the flux deduced from the FOC observation, assuming no red leak
(see \S 3.3). 

The seven archival images which we also include in this atlas were obtained with different
FOC formats, UV filters, and exposure times, providing modified fields of view,
bandpasses, and sensitivities. One of these, of NGC 4151, was obtained in 1995,
after the first {\it HST} servicing mission, and so is free of spherical aberration.
 We elaborate on these differences for each archival exposure
in \S 3.3.

\section{The Atlas}

\subsection{Tables}
The parameters of the 110 observed galaxies are listed in 
Table 1 (southern galaxies) and Table 2 (northern galaxies).
 Following is a column-by-column description of the
table entries, and how they are derived.

(1) NGC designation of the galaxy.
 Footnote ``a'' denotes archival images, generally obtained
with a different FOC format, filter, or exposure time than those of the program galaxies.
See \S 3.3 for details.

(2) UGC designation of the galaxy (northern galaxies only). 

(3) 1950 coordinates, to one-minute accuracy, as listed in the UGC and ESO catalogs. This datum
can be useful for unambiguously identifying the galaxies in these and other catalogs,
since listed coordinates for such large galaxies may vary by arcminutes from catalog to catalog.
Also, some galaxies in the ESO catalog are designated solely by means of these coordinates.

(4)(5) J2000 coordinates of the nucleus, as measured in the STScI GASP system
(see \S 2). These coordinates are generally accurate to $\sim 1''-2''$.
 
(6) $V_h$ -- Heliocentric velocity, as listed in the UGC or ESO catalogs, in km s$^{-1}$.
The selection criterion for inclusion in the sample was $V_h < 2000$ km s$^{-1}$.

(7)(8) $D_a$, $D_b$ -- Major and minor axis diameters, in tenths of arcminutes,
from the UGC and ESO catalogs. The selection criterion for inclusion in the sample
 was $D_a > 60$.

(9) $B_{mag}$ -- Integrated $B$ magnitude, from the UGC and ESO catalogs.

(10) T -- Hubble type, using de Vaucouleurs' T-type classification from the
RC3 catalog (de Vaucouleurs et al. 1991). The correspondence is approximately as follows;
E: $-6$ to $-4$; S0: $-3$ to 0 ; Sa: 1; Sb: 3; Sc: 5; Sd: 7; Irr: 10.

(11) Classif. -- Hubble type and luminosity class using the classification,
when available, 
from the Revised Shapley-Ames Catalog of Bright Galaxies (Sandage \& Tammann 1987).

(12) Sp. -- (northern galaxies only) -- Spectral classification of the nucleus,
 from Ho et al. (1995, 1996a). The designation is as follows.
 L -- LINER (low-ionization nuclear emission-line region);
 H -- H II nucleus; T -- ``transition'' source, between LINER and H~II;
S -- Seyfert nucleus; A -- ``absorption-line'' nucleus with no detected emission lines. A colon
denotes an uncertain classification. The Ho et al. classification
is based on the Filippenko \& Sargent (1985, 1986) optical spectroscopic
survey of the nuclei of a flux-limited ($B < 12.5$ mag) sample of 486 northern galaxies. All but three
of the northern galaxies in the {\it HST} survey (which is diameter- and redshift-limited)
are included in the Filippenko \& Sargent (1985) sample. The effective aperture of the
optical observations is $2''\times 4''$. The Ho et al. (1996a) classification is
assigned after
careful subtraction of absorption-line template spectra, leaving behind only
the emission-line residual;  see Ho et al. (1996a) for further details.

 (13) {\it HST} UT observation date.

 (14) Image rootname in the {\it HST} archive, useful for retrieving the actual
data.

 (15)(16) $f_{UV}, \sigma$ -- Total $f_{\lambda}(2270{\rm~\AA})$ 
in units  of $10^{-15}$ erg s$^{-1}$ cm$^{-2}$ \AA$^{-1}$, integrated 
above the background over the
entire area of the image, and $1\sigma$ uncertainty.
The area of each image is  $22''\times 22''$, except for some of the archival
exposures (marked with footnote ``a'' in column 1),
 which were taken with a different FOC format. See \S 3.3 for details.
 The background was determined as follows.
The mean counts per pixel were measured in seventeen $200\times 200$-pixel squares in the frame, excluding
occulting fingers and distortions in the FOC field (see Baxter et al. 1994), 
and the median  counts per pixel were measured over the entire
exposed part of the frame. The mean of the two lowest among these 18 measurements
was used as the background value, and the standard deviation of the five lowest
among the 18 measurements was used as the uncertainty in the background. The uncertainty in
the background was propagated to an uncertainty on the total net counts in the image.
The count rate above the background was converted to a UV flux density as
described in \S 2. The flux uncertainty due to the background uncertainty was 
combined in quadrature with a 5\% absolute calibration uncertainty (Meurer 1995) to
produce the quoted flux uncertainty. Except for bright and concentrated sources, 
the flux uncertainty is dominated by the uncertainty in the background determination.
The cause of artificial background variations across the image is imperfect
flat-fielding. Furthermore, a systematic error in background determination is
unavoidable due to the small field of view, which covers only a fraction of
the optical extent of these galaxies. Some of this systematic error is accounted
for by the above procedure for estimating the background uncertainty. Nevertheless,
the total UV fluxes quoted here agree well with the 2300 \AA\ fluxes measured for those
galaxies that have also been observed with
 {\it IUE} (Kinney et al. 1993; see also Meurer 1995), which has a comparable
entrance aperture. This
suggests that the regions of the images with the lowest counts are, in fact,
devoid of significant UV emission. The UV fluxes given here should be used with care, and
in conjunction with the UV-morphology classification (column 17) and the visual
appearance of the image. For example, there is low significance to the flux that is
 listed for a galaxy whose image appears blank. The UV fluxes are uncorrected for
Milky Way or external extinction.

(17) Morph. -- UV morphologies roughly describing the {\it HST} image, with the following
symbols. B -- blank image; W -- weak or nearly absent UV emission;
S -- star-forming morphology, with knots and compact sources of UV emission; F -- diffuse,
centrally concentrated emission; P -- unresolved nuclear point-source; R -- circumnuclear
star-forming ring. Some comments on each of these types follow.

We have rechecked the coordinates and pointing of the blank (``B'' morphology)
images, and verified that they are not cases of telescope mispointing.
As a check on the pointing accuracy, there are about 40 images which display a feature
that can be securely associated with the nucleus of the galaxy. In almost
all cases it is within $3''$ of the center of the image, as expected from the
combined uncertainty in the GASP coordinates of the nucleus and the {\it HST} 
pointing accuracy. In the few cases where the nucleus is further from the image center,
this has been traced to inaccurate input coordinates (see \S 3.3). The blank
images are also
not the result of foreground Milky-Way extinction, except for two galaxies, NGC 1560
and NGC 6946, which lie near the Galactic plane. Dust in the disks of the
galaxies themselves
is probably a factor, since 10 out of 13 galaxies with B morphologies have minor-to-major
axis ratios less than 0.5 (i.e., an inclination $>60^{\circ}$). The fact that
the centers of many galaxies are weak UV emitters is confirmed by the detection
of very weak (``W-type'') but significant and centrally concentrated emission 
in many of the galaxies, which establishes that the galactic nucleus is,
indeed, in the field of view.

The compact sources seen in the ``S-type'' morphologies are probably compact 
young  star clusters,
or in some cases individual O and B stars. Similar objects have been detected with {\it HST} 
in a variety of starburst environments (e.g., Meurer et al. 1995; Maoz et al. 1996).
They will be studied in further detail by Ho et al. (1996b).

 The diffuse ``F-type'' emission
occurs in some of the early-type spirals and the ellipticals in the sample. We believe
that, in general, this observed feature is dominated by
 actual UV emission from an evolved spheroidal stellar population (the
``UV-upturn''; see, e.g., Burstein et al. 1988),
 rather than visual-band emission leaking through the F220W filter, based
on several tests described in \S 2. In individual cases, however, confirmation by means of  
blue and near-UV imaging photometry of the center of each of these galaxies is required.

 Galaxies with bright nuclear UV point sources (``P-type'') have been discussed
in detail by Maoz et al. (1995), especially in the context of low-luminosity AGNs.
They showed that $\sim 20\%$ of the northern LINER galaxies display a nuclear point source
in the FOC images,
with a UV flux that, if extrapolated beyond the Lyman limit,
 could be sufficient to produce the observed strength of optical
emission lines through photoionization. While this was the first direct detection
of what may be the AGN-like ionizing source in LINER galaxies, it raised the question
of why such a source is {\it not} detected in 80\% of LINERs.
Several of the P-type  sources in in this paper,
 especially the weak ones and those in archival
images, were not included in Maoz et al. (1995). However, the fraction of ``UV-bright''
LINERs, or LINERs plus Seyferts, remains unchanged. For example, among the 35 northern
galaxies with spectral classification T, L, or S (transition-type, LINER, or Seyfert),
nine have a nuclear UV point source.

 The five circumnuclear rings
in the sample (designated ``R'') have been discussed in detail by Maoz et al. (1996),
who showed that a large, possibly dominant fraction of the UV light in these objects
is emitted by the numerous compact sources distributed along the rings.
These sources are probably young and massive star clusters that will remain bound,
similar to those seen with {\it HST} in other starburst environments.

\subsection{Pictorial Atlas}

Figures 1 and 2 are grey-scale representations of the FOC images of the
southern and northern galaxies, respectively.
Each of the six frames on each page has a scale of
 $22''\times 22''$, and has been rotated so that north is up
and east is to the left. Archival exposures
are marked with an asterisk next to the galaxy name, and
in some cases have a different field of view and sensitivity;
 see \S 3.3
for details. The grey scale of each image is chosen to bring out the
most interesting details. However, these 
snapshots cannot convey all of the useful information in many of the images,
and the reader is advised to refer to \S 3.3,
or to retrieve the actual data from the {\it HST} archive,
if complete details are required. This is true both for bright
sources, where a single grey scale cannot show details of different contrast,
and for weak sources (``W-type'' morphology), where the faint emission
is lost in the reproduction.

\subsection{Notes on Individual Objects}
\subsubsection{Southern Galaxies}
 \indent NGC 247 (0044$-$2102).--
This late-type galaxy has 
a bright UV point source at a position consistent
with the optical nucleus of the galaxy, which has an H II-like spectrum
(Maoz et al. 1995). In addition
there are numerous compact sources in the image. See Maoz et al. (1995)
for further details.
\\ \indent NGC 300 (0052$-$3757).--
There are about a dozen compact sources in the UV image,
the brightest of which is resolved (FWHM$\approx 0.2''$)
and is probably at the nucleus of this face-on
Sd galaxy. The second brightest source, $3''$ to the east, 
is unresolved (FWHM$< 0.1''$).
\\ \indent NGC 1079 (0241$-$2912).-- 
The small ($4''$ diameter) circumnuclear star-forming ring 
seen in the {\it HST} image of this S0 galaxy was discovered
by this survey. A detailed study, including rotation curves
along the major axis and photometry of the compact young
star clusters that contribute much of the UV light, can 
be found in
 Maoz et al. (1996).
\\ \indent NGC 1291 (0315$-$4117).--
The central peak of the diffuse UV-light distribution
may be unresolved in this SBa galaxy.
\\ \indent NGC 1332 (0324$-$2130).--
This is an inclined S0 galaxy.
A thin, unresolved dust lane crosses 
the diffuse light distribution about
$0.3''$ northeast of the central peak from south-east
to north-west, along the orientation of the galaxy's
major axis.
\\ \indent NGC 1385 (0335$-$2440).--
This is a disturbed-looking Scd galaxy.
The field of view includes part of some large structure,
possibly the western spiral arm of the galaxy,
that is bright in the UV and includes compact sources
and diffuse emission. The UV-dark northern portion of the
image is at the end of a large dusty spiral arm that crosses
the northern part of the galaxy in optical images. The nuclear 
position is difficult to determine in optical images,
and the bright knot on the western side of the UV image could be
the galaxy nucleus.
\\ \indent NGC 1433 (0340$-$4722).--
The circumnuclear star-forming ring seen in the UV image of this
SBb galaxy is studied in detail
by Maoz et al. (1996).
\\ \indent NGC 1512 (0402$-$4329).--
The circumnuclear star-forming ring seen in the UV image of this
SBb galaxy is studied in detail
by Maoz et al. (1996); $\sim 40\%$ of the UV light comes from
compact young star clusters distributed along the ring. About 25\%
of the UV light is from the single bright ``super star cluster''
on the south-east side, which has an observed UV luminosity 
$L_{\lambda} (2300{\rm  \AA})\approx 10^{37}$ erg s$^{-1}$  \AA$^{-1}$
and a radius of only 2--3 pc.
\\ \indent NGC 1543 (0411$-$5751).--
The UV image of this SB0/a galaxy shows a
 centrally-peaked diffuse light distribution with an
unresolved core or a faint central point source.
\\ \indent        ESO 0450$-$2519.--
The single weak
 source detected in this diffuse late-type spiral galaxy
is unresolved or marginally resolved.
\\ \indent NGC 2997 (0943$-$3057).--
This circumnuclear star-forming ring is studied in detail
by Maoz et al. (1996). Utilizing also a $V$-band
{\it HST} exposure of this galaxy, they show that 
the numerous compact (few pc radius) sources distributed along the ring
are most probably young ($<100$ Myr) and massive ($\sim 10^5 M_{\odot}$)
clusters of stars that are gravitationally bound. These
clusters, which appear similar to those in the many galaxies with  ``S-type''
UV morphology presented here, may therefore evolve into objects
similar to globular clusters.
\\ \indent NGC 4976 (1305$-$4914).--
The UV image of this bright S0 galaxy shows only a
faint, centrally-peaked diffuse light distribution with an
unresolved core or a weak central point source.
\\ \indent NGC 5084 (1317$-$2133).--
The UV image of this S0 galaxy shows only a
faint, centrally-peaked diffuse light distribution with an
unresolved core or a very weak central point source, with flux of order
a few $10^{-18}$ erg s$^{-1}$ cm$^{-2}$ \AA$^{-1}$.
\\ \indent NGC 5102 (1319$-$3622).--
The archival UV image of this nuclear-starburst S0 galaxy
(e.g., Bica \& Alloin 1987) is a 1137~s F220W
exposure with the $512\times512$ pixel $f/96$ FOC
format, which gives a field of view of $10''\times 10''$.
A bright, unresolved central point source is
evidenced by the saturated central pixels and
the sharp diffraction rings, surrounded by a 
high surface-brightness elliptical light distribution
with major axis oriented north-east to south-west.
A linear structure, probably an edge-on disk,
  can be traced out to about $1''$
from the nucleus on both sides along the same
direction.
\\ \indent NGC 5253 (1337$-$3123).--
This nearby (4.1 Mpc; Sandage et al. 1994) starburst galaxy has been studied from
radio to X-ray wavelengths, including {\it IUE} UV spectroscopy,
which has revealed Wolf-Rayet spectral features (see Kinney
et al. 1993, and references therein, for a summary).
The FOC image presented here has been extensively
 studied by Meurer et al. (1995).
This archival {\it HST} image is a 500 s
exposure with the F220W filter plus a F1ND neutral-density
filter, which provides a factor 2.5 attenuation in flux.
The exposure is therefore 1/3 as deep as our standard-setup exposures.
Nevertheless, the image shows  numerous bright sources.
As shown by Meurer et al. (1995), these are
compact young star clusters of the type detected
with {\it HST} in other starburst environments, as well as 
individual stars. This galaxy
has the highest total UV flux in the sample.
\\ \indent        ESO 2159$-$5132.--
The image of this nearby irregular galaxy 
shows a large number of compact sources, most of which are 
probably individual O and B stars.
\\ \indent NGC 7462 (2260$-$4106).--
The compact sources and diffuse emission in this nearly edge-on Sc galaxy
are concentrated in a narrow strip running approximately east-west,
along the galaxy's major axis.
\\ \indent        ESO 2333$-$3903.--
This faint but strongly-nucleated Sd galaxy has an unresolved
(FWHM $<0.1''$)  and weak  central UV source, with a 2300 \AA\
 flux $\sim 1\times
10^{-16}$ erg s$^{-1}$ cm$^{-2}$ \AA$^{-1}$.

\subsubsection{Northern Galaxies}
 \indent M31 (NGC 224,  UGC 454,  0040+4100).--
This archival image of the center of the Andromeda galaxy is
one of three exposures described in detail by King et al. (1992)
and King, Stanford, \& Crane (1995). It is a 
1677 s exposure with the $f/48$ FOC in its $512\times 1024$
``zoomed'' pixel mode, providing a $44''\times 44''$ field of view.
The filter is F175W, whose peak transmission is at $\sim 1900$ \AA. 
Considering the response of this setup and the exposure time, the image
is similar in sensitivity to the standard setup. The image shows bright,
centrally-peaked diffuse emission that is asymmetric at its peak.
King et al. (1995) show that this peak coincides with the fainter of 
two peaks seen in optical {\it HST} images, and with
the dynamical center of the galaxy (Lauer et al. 1993). 
Many individual stars are also visible, and are discussed in detail 
by King et al. (1992).
\\ \indent NGC 404  (UGC 718,  0106+3527).--
This S0 galaxy contains one of the UV-bright LINERs studied in detail
by Maoz et al. (1995), showing a prominent unresolved nuclear
point source.
\\ \indent NGC 672  (UGC 1256, 0145+2711).--
The image of this Scd galaxy shows a resolved (FWHM$\approx0.2''$) central
source, plus several additional faint compact sources.
\\ \indent NGC 1023 (UGC 2154, 0237+3850).--
This is an SB0 galaxy that has also been observed
with {\it IUE} (Kinney et al. 1993). The integrated flux
from the weak, centrally-concentrated diffuse emission
seen in the {\it HST} image agrees well with the
{\it IUE} 2300 \AA\ flux. This galaxy has also been imaged
with {\it HST} by Lauer et al. (1995) in the $V$ band, where
it displays an unusually high central surface brightness. After deconvolution
with the spherically-aberrated PSF, they estimate a surface
brightness of 12.7 mag arcsec$^{-2}$ in the central $0.022''$, and 
13.8 mag arcsec$^{-2}$ at the $0.1''$ radius. As in the other 45
early-type galaxies they 
have studied, the brightness profile continues to increase
with decreasing radius
all the way down to the {\it HST} angular resolution limit.
The weak emission we detect from this galaxy in the UV, compared with the high
visual surface brightness, confirms that the red leak through
the F220W filter makes a negligible contribution to the counts
we detect in the UV in other objects (see \S 2, and notes on NGC  4636, below).
\\ \indent NGC 1560 (UGC 3060, 0427+7146).-- The absence of UV flux from
this galaxy may be due in part to its low Galactic latitude
 ($+16^{\circ}$), at which a factor $\sim 5$ extinction in 2300 \AA\ flux
is expected (Burstein \& Heiles 1982).
\\ \indent NGC 2903 (UGC 5079, 0929+2143).--
This ``hot-spot'' Sc galaxy has been observed
with {\it IUE}, and its spectrum indicates the
presence of a mixture of early-type and late-type stars
(Kinney et al. 1993).
The {\it HST} image shows many bright compact sources, presumably
young star clusters.
\\ \indent NGC 3079 (UGC 5387, 0958+5555).--
There is very little UV emission detected in the image
of this LINER galaxy. There is a hint of a dust lane
crossing the image from north to south, at an orientation
matching that of the galaxy's major axis.
\\ \indent NGC 3319 (UGC 5789, 1036+4156).--
The elongated diffuse emission in the UV image is aligned
with the strong bar of this Scd galaxy. There is an
unresolved (FWHM$< 0.1''$) nuclear source with $f_{\lambda}(2300{\rm~\AA})=
1.4\times 10^{-16}$ erg s$^{-1}$ cm$^{-2}$ \AA$^{-1}$,
 and several faint compact sources.
\\ \indent NGC 3344 (UGC 5840, 1040+2511).--
This starburst nucleus has a bright unresolved nuclear
point source and has been studied in more detail 
by Maoz et al. (1995).
\\ \indent NGC 3368 (UGC 5882, 1044+1205).--
The UV image of this bright Sab LINER galaxy
 has a single, unresolved compact source 
with $f_{\lambda}(2300{\rm~\AA})\sim
5\times 10^{-17}$ erg s$^{-1}$ cm$^{-2}$ \AA$^{-1}$
about $0.1''$ north of the centroid of a weak diffuse
light distribution.
We have verified by means of a ground-based CCD image
that the source is at the galaxy's nucleus position.
It is $5.5''$ south of the center of the image because
 the galaxy nucleus coordinates
input to {\it HST} were off by that much due to galaxy isophote
distortion by a dust lane in the GASP image.
This is another example of
a ``UV-bright'' LINER (Maoz et al. 1995), though with
a central point source fainter by 1-2 orders of magnitude.
In contrast to the other UV-bright LINERs, whose observed
UV fluxes are sufficient to explain their observed emission-line
fluxes by photoionization, in this source the observed
UV flux is 20 times less than the minimum required (see Maoz 1996).
 This LINER
is therefore reddened or not excited through photoionization, and
could be an intermediate case between UV-bright and UV-dark.
\\ \indent NGC 3486 (UGC 6079, 1057+2914).--
There is a weak ($f_{\lambda}(2300{\rm~\AA})=
1\times 10^{-16}$ erg s$^{-1}$ cm$^{-2}$ \AA$^{-1}$), unresolved
(FWHM$<0.1''$) point source in the nucleus
of this galaxy, which is a LINER/weak-Seyfert 2 (Ho et al. 1996a).
The H$\alpha$ flux measured by Ho et al. (1996a) is about 
$1\times 10^{-14}$ erg s$^{-1}$ cm$^{-2}$. Using the argument of Maoz et al. (1995)
and Maoz (1996) in the cases of other LINERs in the sample, a power-law
 extrapolation beyond the Lyman limit of the observed UV flux provides
enough ionizing photons to explain the observed H$\alpha$ flux through
photoionization in this object as well. The central point source could
therefore be the AGN continuum source.
\\ \indent NGC 3521 (UGC 6150, 1103+0014).--
The central diffuse light distribution in the UV image of this
Sbc galaxy has a double-peaked appearance, apparently because
it is crossed by an obscuring dust feature. The feature is
reminiscent of the X-shaped dust lanes in M51 (NGC  5194; see below),
but this requires confirmation with data of higher signal-to-noise ratio (S/N).
\\ \indent NGC 3627 (UGC 6346, 1117+1315).-- The nucleus of this
bright Sb galaxy has a prominent optical
emission line spectrum that is intermediate between that of a LINER and 
and an H II nucleus (Ho et al. 1996a).
 The UV image shows a weak, diffuse centrally-peaked
light distribution. It is offset from the image center due to an inaccuracy in the
nuclear position that was estimated from a GASP  image and input to {\it HST}.
There is a hint of spiral structure which, however, requires confirmation.
\\ \indent NGC 3718 (UGC 6524, 1129+5320).-- This Sa galaxy has a prominent
optical LINER spectrum, including a broad H$\alpha$ component similar to
that of Seyfert 1 galaxies (Ho et al. 1996a).
Its UV image is blank, and this may be related  to the fact that the optical line ratios
are indicative of reddening. A dust lane that crosses
optical images of the galaxy near the nucleus is a candidate agent of these effects. 
\\ \indent NGC 4151 (UGC 7166, 1208+3941).--
This is probably the best studied Seyfert 1 galaxy.
The archival image is a 900 s exposure with the $f/48$ FOC in its $512\times 1024$
``zoomed'' pixel mode. The image was taken after the {\it HST} refurbishment mission
and the installation of COSTAR, and so is free of spherical aberration.
The field of view is $28''\times 28''$, with a de-zoomed pixel size of $0.014''$.
 The image was obtained with a
combination of the F220W and F275W filters, giving an effective
peak transmission at $\sim 2450$ \AA. 
Considering the response of this setup and the exposure time, the image
has 1/4 the sensitivity of the standard setup for diffuse sources,
 but better sensitivity
to compact sources because of the improved optics. The bright Seyfert 1
nucleus is strongly saturated. Its brightness cannot be reconstructed,
as in Maoz et al. (1995), due to the absence of the spherical-aberration
diffraction rings. In addition to the nuclear point source, there is
a diffuse-light distribution centered $2''$ south of the nucleus.
Excluding the counts from the central source, the
flux from the diffuse component is $1.2\times 10^{-13}$
erg s$^{-1}$ cm$^{-2}$ \AA$^{-1}$. This value casts doubt on the component's reality,
since measurements with {\it IUE}, whose aperture would include much of
this extended (and presumably non-variable) emission,
 have been at times up to a factor of 6 lower at this wavelength
(e.g., Ulrich et al. 1991).
The diffuse component may be an artifact of the poor quality of
the $f/48$ FOC flat-field, or of a dark current in this camera,
which has had operational problems since December 1992. 
The large arc to the west is an artifact.
\\ \indent NGC 4192 (UGC 7231, 1211+1510).-- This highly-inclined Sb galaxy has
a nucleus with strong
optical emission lines, with line ratios intermediate between a LINER
and an H II nucleus, and indicative of reddening (Ho et al. 1996a).
Its UV image shows three amorphous knots of very weak emission.
\\ \indent NGC 4214 (UGC 7278, 1213+3637).--
This starburst Magellanic irregular galaxy has been studied
in the UV with {\it IUE} (Huchra et al. 1983; Hartmann, Geller, \& Huchra
1986), and in the optical (Sargent \& Filippenko 1991),
and shows Wolf-Rayet signatures.
The archival {\it HST} FOC image shown here is analyzed, along with
{\it HST} spectroscopy, by Leitherer et al. (1996).
It is a 1200 s exposure with the standard format,
 but with a F2ND
neutral density filter, which attenuates the flux by a factor of 5,
 in addition to the F220W filter. The sensitivity is then
0.4 times that of the standard setup. The image shows a bright
central point source, surrounded by numerous compact sources.
Most of the total UV flux, which is
in excellent agreement with that measured by {\it IUE} at 2300 \AA\
(Kinney et al. 1993), comes from the central source. Leitherer et al. (1996)
argue that most of the fainter sources are individual stars. This galaxy
has the second largest total UV flux in the sample (after NGC  5253).
\\ \indent NGC 4438 (UGC 7574, 1225+1317).-- The UV image of this
S0 galaxy shows only a weak amorphous patch of UV emission, about $5''$ in size.
Filippenko \& Sargent (1985) note that the strong LINER-like emission that is seen
in optical spectra comes from an extended region.
\\ \indent M87 (NGC 4486, UGC 7654, 1228+1240).--
This is the central elliptical galaxy in the Virgo cluster
and a much-studied AGN, with a LINER nucleus and a jet that is visible
at radio, optical, and UV wavelengths.
UV images of the nucleus and the jet have been previously
published by Boksenberg et al. (1992).
This archival image is a 1200 s exposure with the FOC/96 in its 
``normal'' (i.e., unzoomed) $512\times 512$ pixel format, giving
a $10''\times10''$ field of view, and the standard F220W filter.
The exposure is 1.6 times more sensitive than the standard setup.
The image shows a bright central point source, and the
knots of the UV jet emerging westward. The UV flux from the central
point source, $1\times 10^{-15}$ erg s$^{-1}$ cm$^{-2}$ \AA$^{-1}$,
 is sufficient to provide the observed H$\alpha$ flux
through photoionization, as in the other UV-bright LINERs in the
sample (Maoz et al. 1995; Maoz 1996).
\\ \indent NGC 4569 (UGC 7786, 1234+1326).--
This Sab galaxy contains one of the UV-bright LINERs studied in more detail
by Maoz et al. (1995). Apart from the bright unresolved nuclear
point source, there is some faint extended emission $0.65''$ south of the 
nucleus.
\\ \indent NGC 4579 (UGC 7796, 1235+1205).--
This is one of the UV-bright LINERs studied in more detail
by Maoz et al. (1995), showing a bright unresolved nuclear
point source. The main source is surrounded by several
fainter sources, giving the diffraction rings in the image their
asymmetric appearance.
\\ \indent NGC 4636 (UGC 7878, 1240+0257).-- 
This E/S0 galaxy has an optical LINER  spectrum, 
with broad H$\alpha$ wings (Ho et al. 1996a).
The UV image shows diffuse, centrally-concentrated emission. This galaxy has also
been imaged with {\it HST} by Lauer et al. (1995) in the $V$ band,
where it shows similar structure. The $V$-band brightness profile
increases with decreasing radius down to the {\it HST} resolution, though with a
shallower slope than most of the other early-type
 galaxies studied by Lauer et al. (1995). They estimate (after deconvolution
with the spherically-aberrated PSF) a surface
brightness of 16.5 mag arcsec$^{-2}$ in the central $0.022''$, and 
16.7 mag arcsec$^{-2}$ at the $0.1''$ radius. The comparable UV brightnesses
of this galaxy and NGC  1023 (which was also imaged by Lauer et al. 1995),
 despite the much higher optical surface brightness
of NGC  1023, confirms that the red leak through the F220W filter contributes
negligibly to the counts in the FOC exposures, and that most of the signal
detected in the FOC exposures is, in fact, the result of UV flux (see \S 2.2 and
notes on NGC  1023).
\\ \indent NGC 4725 (UGC 7989, 1248+2546).--
This bright SBb galaxy hosts a weak Seyfert 2 nucleus
(Ho et al. 1996a). The UV image shows a diffuse centrally-peaked
light distribution, with a marginally resolved core of width
$\approx 0.2''$.
\\ \indent NGC 4736 (UGC 7996, 1248+4123).--
This ringed Sab galaxy has many peculiar features, suggesting it
is in the final stages of a merger. It
is one of the UV-bright LINERs studied in more detail
by Maoz et al. (1995), and has two bright unresolved
UV point sources, one on the nucleus and one $2.5''$ to the
north. Using the algorithm of Maoz et al. (1996), we measure
for the nuclear point source a UV flux density
$f_{\lambda}(2300{\rm~\AA})=
1.9\times 10^{-16}$ erg s$^{-1}$ cm$^{-2}$ \AA$^{-1}$,
and for the northern source
$f_{\lambda}(2300{\rm~\AA})=
2.6\times 10^{-16}$ erg s$^{-1}$ cm$^{-2}$ \AA$^{-1}$.
 The nuclear point source sits on top of a high 
surface-brightness diffuse component that extends into a number of
concentric arcs. Several fainter compact sources are also 
visible.
\\ \indent NGC 4866 (UGC 8102, 1256+1426).--
The UV image of this Sa galaxy, whose nucleus has a LINER spectrum
(Ho et al. 1996a), shows a weak, centrally-peaked diffuse light
distribution. There is a hint of a dust lane crossing the nucleus
in the UV image,
but this needs confirmation with higher S/N.
\\ \indent NGC 5005 (UGC 8256, 1308+3719).--
This is a bright Sbc galaxy with a LINER nucleus (Ho et al . 1996a).
We have compared the gross features of the UV image to a narrow-band
ground-based H$\alpha$ image, and verified that the faint diffuse emission
 that appears several
arcseconds west of the center in the UV image is in fact the nucleus
of the galaxy. The knot of UV sources on the south-east edge of the image
corresponds in the ground-based image to an H$\alpha$ knot. 
There may be a very faint compact V-shaped source of
 UV emission on top of the diffuse nuclear emission, but this requires
confirmation at higher S/N.
\\ \indent NGC 5055 (UGC 8334, 1313+4217).--
This is one of the UV-bright LINERs studied in more detail
by Maoz et al. (1995), showing a bright marginally-resolved nuclear
point source. Many fainter compact sources are also
visible in the northern and southern corners of the image.
\\ \indent M51 (NGC 5194, UGC 8493, 1327+4727).--
This face-on Sbc Seyfert 2 galaxy forms an interacting pair with NGC  5195
and has been the focus of many studies.
The archival {\it HST} image is a 2277 s exposure with the $f/96$ FOC in its 
``normal'' (i.e., unzoomed) $512\times 512$ pixel format, giving
a $10''\times10''$ field of view, and the F275W filter, at an 
effective wavelength $\sim 2800$ \AA. The longer wavelength
and exposure time make this image about 8 times more sensitive
than the standard setup. The image shows X-shaped
dust lanes crossing the bright diffuse light distribution, previously
noted in visual-band {\it HST} images (Ford et al. 1992; see Maran \& Kinney 1993),
and a number of compact sources on the outskirts of the field.
\\ \indent NGC 5195 (UGC 8494, 1327+4731).--
This is the companion galaxy interacting with M51 (NGC  5194). The UV image
is nearly blank, with only some faint diffuse emission.
\\ \indent NGC 5248 (UGC 8616, 1335+0908).--
This Sbc galaxy contains one of the circumnuclear star-forming rings studied in detail
by Maoz et al. (1996).
\\ \indent NGC 5322 (UGC 8745, 1347+6026).-- This is an E4 elliptical galaxy with 
a weak LINER nucleus (Ho et al. 1996a). The UV image shows weak, diffuse, centrally-concentrated
structure that is elongated in the east-west direction, like the galaxy's major axis
in optical images. A pre-COSTAR {\it HST} $V$-band image of the galaxy
(Lauer et al. 1995) shows similar structure, and also a thin dust lane crossing
the nucleus. There is a hint of this dust lane, also in the east-west direction,
in the UV image.
\\ \indent NGC 5474 (UGC 9013,  1403+5354).--
This Scd galaxy is one of a group that is tidally interacting
with the large galaxy M101 (NGC  5457; Kinney et al. 1993).
 The image shows many
compact sources, presumably young star clusters and individual O and B stars.
\\ \indent NGC 5866 (UGC 9723,  1505+5557).-- The UV image of this S0 galaxy
displays weak, amorphous emission. The optical spectrum of the nucleus 
is a ``transition type'' between LINER and H II, with weak broad H$\alpha$ wings
(Ho et al. 1996a).
\\ \indent NGC 6946 (UGC 11597, 2033+5959).-- This is a well-known
low-surface-brightness, face-on,
 Scd starburst galaxy (e.g., DeGioia-Eastwood 1985; 
Devereux \& Young 1993). The absence of UV flux in the FOC exposure
may be due in part to its low Galactic latitude
 ($+12^{\circ}$), at which a factor $\sim 5$ extinction in 2300 \AA\ flux
is expected (Burstein \& Heiles 1982).
\\ \indent NGC 7331 (UGC 12113, 2234+3409).-- This bright Sb galaxy 
has the optical spectrum of a ``transition object'' between a LINER
and an H II nucleus (Ho et al. 1996a). The UV image shows a faint
but strongly peaked diffuse light ditribution, and some additional emission
on the eastern edge of the field.

\section{Summary}

We have presented {\it HST} UV images of the central regions
 of 110 large nearby galaxies.
   The observed galaxies are an unbiased selection from
  a complete sample of all large and nearby
 galaxies. This is the first such UV imaging survey 
of normal galaxies.
   The images display an 
   assortment of morphologies and UV brightnesses. These include bright
 unresolved ($<0.1''$, usually corresponding to a few pc)
 nuclear point sources, compact young star clusters scattered in the field or
arranged in circumnuclear rings, centrally-peaked diffuse light distributions,
and galaxies with weak or undetected UV emission. For every galaxy,
we have measured the integrated
$\sim 2300$ \AA\ flux, classified the UV morphology, and tabulated these
parameters along with others from the literature.

 From intercomparing the galaxy parameters, the following trends emerge.
As seen in Figure 3, the centrally-peaked diffuse (F-type) UV morphology tends to occur in
spirals earlier than type Sc, and in S0s and ellipticals. This UV morphology
is therefore the signature of a spheroidal population, although not all
spheroids produce it. Galaxies with this morphology tend to
have a lower UV flux, integrated over the central $20''$, than others.
The central UV flux of the F-types
 is not correlated with the integrated $B$ magnitude of the
entire galaxy. However, there are no galaxies in the sample fainter
than $B\approx 12$ mag with an F-type morphology (or, for that matter,
with a nuclear point source [P]). The latter three properties can be
seen in Figure 4.

 Conversely (see Figure 3), the
star-forming (S-type) UV morphology tends to occur more in late-type galaxies,
but it can be found in all Hubble types. The highest integrated central UV fluxes occur
 in galaxies with S-type morphology although, again, a large range in flux
exists. In these S-types,
 the central UV flux is correlated with the galaxy's integrated $B$ magnitude,
but with a large scatter. These trends are shown in Figure 4.
 Galaxies that are null (B) or weak (W) UV emitters
do not have significantly fainter integrated $B$ magnitudes than galaxies
that are UV-bright. As already noted, the fraction of galaxies with B-type
morphology that have an inclination $>60^{\circ}$ is  10/13, as opposed
to the 1/2 expected from a random ditribution (and which the other types
obey). This suggests that dust in the disk of a nearly edge-on galaxy sometimes
stifles completely what may have been a W-type emitter. On the other hand,
the converse is not true; we find no trend of UV flux with axis ratio, and
 many highly-inclined galaxies are plentiful UV emitters.

Comparing the northern sample's UV properties to the optical spectral classification
of Ho et al. (1996a), we find that galaxies with a star-forming (S) UV morphology
never have a ``pure'' LINER-type (L) spectrum. This is as expected, since the UV sources
in the S-types are probably young star clusters or individual O and B stars, 
whose ionization will add H II-like spectral features to a LINER spectrum.
Also, galaxies with a nuclear point source are never devoid of emission lines
in the optical spectrum (spectral classification ``A'')
 suggesting that, if a nuclear UV continuum source is present,
there is usually enough gas in the nucleus to be ionized by it and produce emission lines.
As shown in Figure 5, the highest central UV fluxes are measured in galaxies with
 an H II-like (``H-class'')
optical spectrum, and the median UV flux of this class is several times higher 
than that of galaxies with the other types of nuclear spectra.
 This is also not surprising, since Ho et al. (1996a) show that
such spectra tend to occur in late-type galaxies, and we have found (see above)
that these are the Hubble types with the highest UV fluxes.

 The data we have presented can be useful for studying star formation
and low-level nuclear activity in galaxies, for identifying 
galaxies with significant ``UV-upturn'' emission from an evolved stellar
population, and for searching for interconnections between these
phenomena. These are the subjects of recent
and forthcoming publications based on this survey.

\acknowledgements
We are grateful to J. N. Bahcall for his contribution to the earlier
stages of this survey, to T. Lauer for providing
information on the central surface brightness of two early-type galaxies,
and to the referee, A. Sandage, for constructive comments.
This work was supported by grant GO-3519
 from the Space Telescope
Science Institute, which is operated by AURA, Inc., under NASA
contract NAS 5-26555. A. V. F. and L.~C.~H. also acknowledge
support by grant AR-4911 from the Space Telescope
Science Institute.
D. M. and H. -W. R acknowledge support by  the U.S.-Israel
Binational Science Foundation grant 94-00300.

\newpage
\figcaption{ 
Grey-scale representations of the FOC images of the
southern galaxies.
Each of the six frames on each page has a scale of
 $22''\times 22''$, and has been rotated so that north is up
and east is to the left. Archival exposures
are marked with an asterisk next to the galaxy name, and
in some cases have a different field of view and sensitivity;
 see \S 3.3 
for details. The grey scale of each image is chosen to bring out the
most interesting details. However, these 
snapshots cannot convey all of the useful information in many of the images,
and the reader is advised to refer to \S 3.3,
or to retrieve the actual data from the {\it HST} archive,
if complete details are required.}

\figcaption{ Same as Figure 1, for the northern galaxies.}

\begin{figure}
\figurenum{3}
\epsscale{1.4}
\plotone{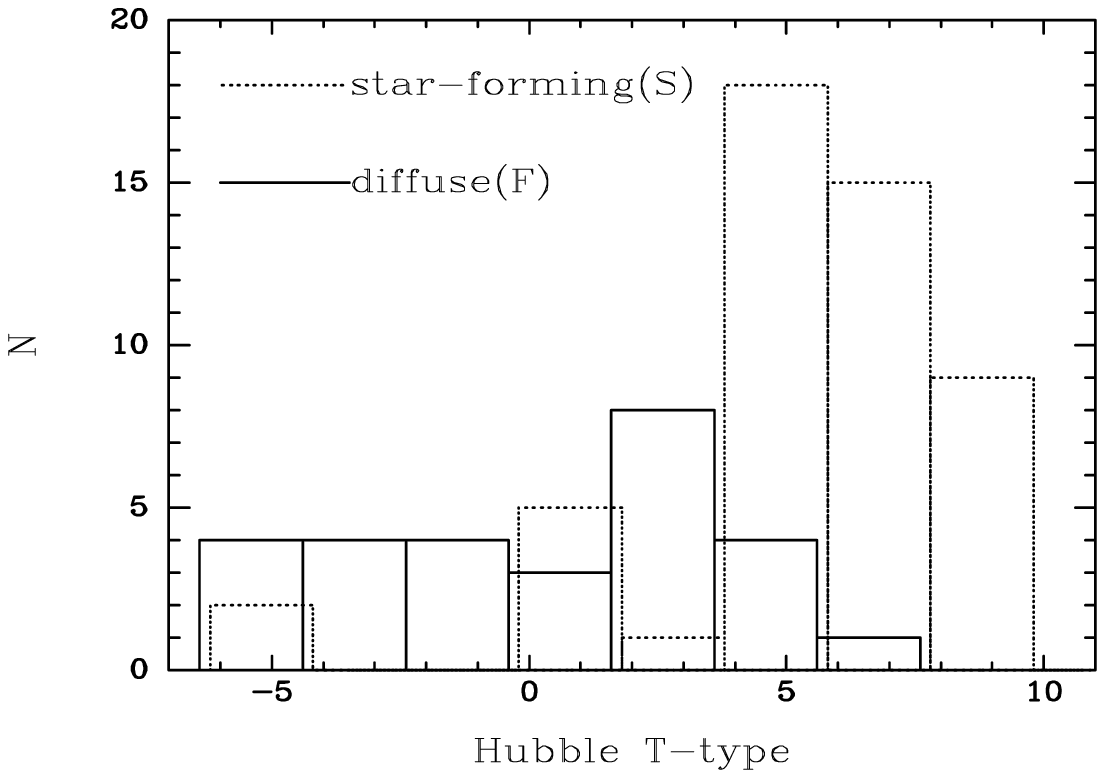}
\caption{ The distribution of galaxy Hubble types (labeled
according to de Vaucouleurs's T-types) for galaxies with
a star-forming (``S-type''; dotted histogram) UV morphology,
and galaxies with a diffuse centrally-peaked (``F-type''; solid
histogram) UV morphology. F-types tend to occur in earlier-type
galaxies, and are apparently associated with a galaxy spheroid.
S-type UV morphologies are found preferentially in the centers
of late-type spirals, but can occur in all Hubble types.}
\end{figure}

\begin{figure}
\figurenum{4}
\epsscale{1.4}
\plotone{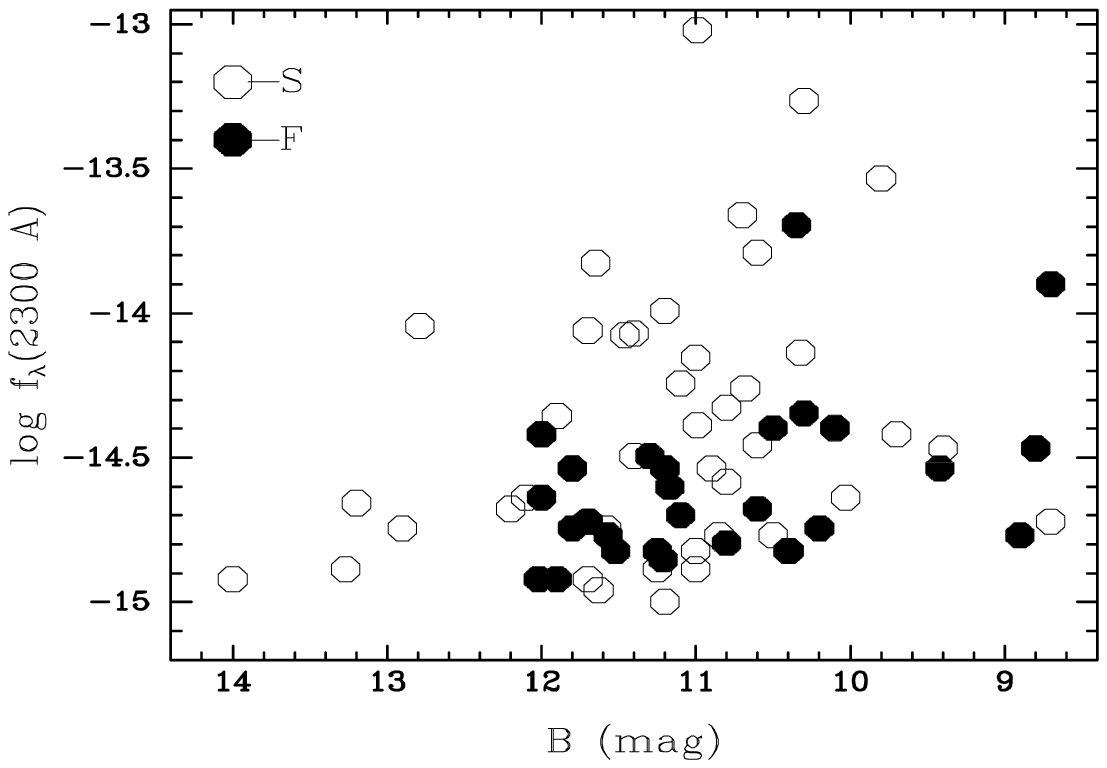}
\caption{ Observed flux density at 2300 \AA, integrated over
the central $22''\times 22''$ of each galaxy, vs. $B$ magnitude
integrated over the whole galaxy, for galaxies with F-type
UV morphologies (filled circles) and S-type UV morphologies (empty circles).
The central UV flux of F-types is uncorrelated with the galaxy's
integrated $B$ magnitude, but such  a morphology is found in
the sample only in galaxies brighter than $B=12$ mag.
For S-types, on the other hand, the central UV flux appears to be
loosely correlated with the total $B$ magnitude.
The typical UV flux from F-types is several times lower than that 
of S-types, which have the highest UV fluxes in the sample.}
\end{figure}

\begin{figure}
\figurenum{5}
\epsscale{1.4}
\plotone{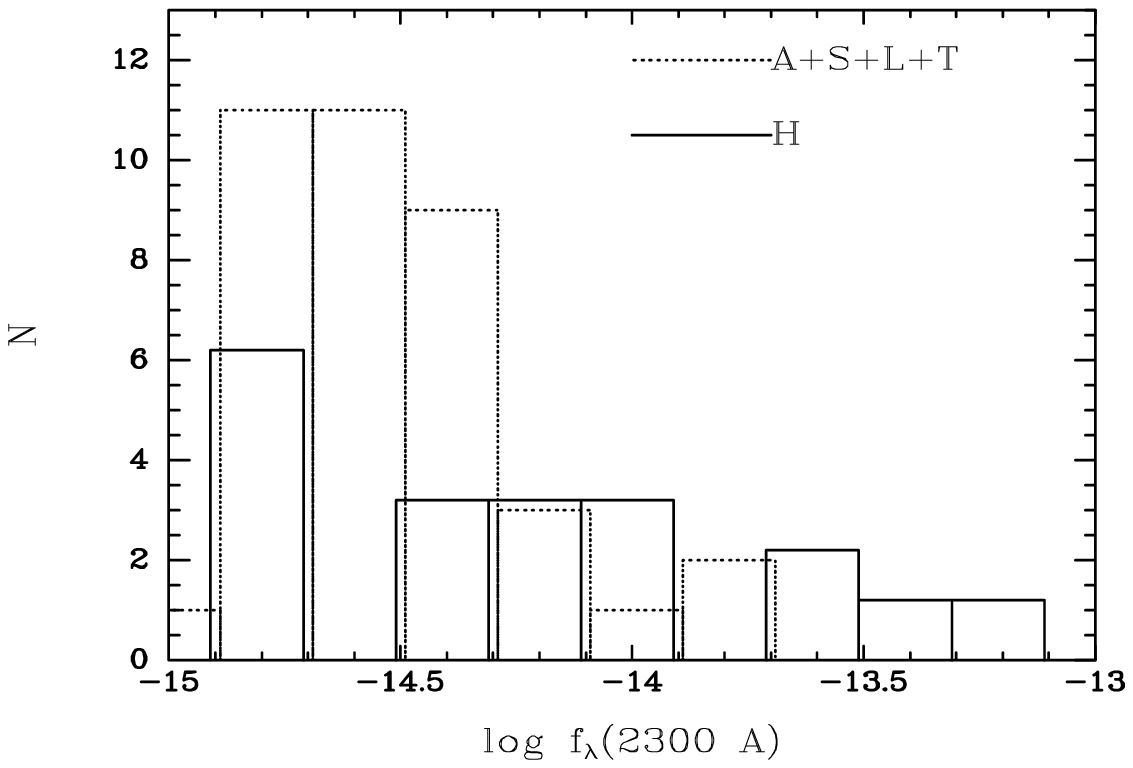}
\caption{ The distribution of central UV flux density, as observed
with the FOC, for galaxies with
an H II-like optical nuclear spectrum (``H-type'' spectral classification;
 solid histogram),
and galaxies with other optical nuclear classifications (Absorption (A), Seyfert (S),
LINER (L), or transition-type (S); dotted
histogram). Galaxies with H II-like optical nuclear spectra tend to have higher 
integrated central UV flux, and the galaxies with the highest UV fluxes are of 
this type.}
\end{figure}

\end{document}